\documentclass{ws-ijmpa}
\usepackage[super,compress]{cite}
\usepackage{graphicx}
\begin{document}
\markboth{Authors' Names}{Noncommutative Brownian motion}

%
\catchline{}{}{}{}{}
%

\title{Noncommutative Brownian motion}

\author{Willien O. Santos
}

\address{Colegiado de Fisica, Universidade Federal do Reconcavo
da Bahia\\
45300-000, Amargosa, BA, Brazil\\
\_willien@ufrb.edu.br\_}

\author{Guilherme M. A. Almeida}

\address{Instituto de Fisica, Universidade Federal de Alagoas\\
57072-900, Maceio, AL, Brazil
}
\address{Departamento de Fisica, Universidade Federal de Minas Gerais\\
Caixa Postal 701, 30161-970, Belo Horizonte, MG, Brazil\\
\_gmaalmeidaphys@gmail.com\_}

\author{Andre M. C. Souza
}

\address{Departamento de Fisica, Universidade Federal de Sergipe\\
49100-000, Sao Cristovao, SE, Brazil\\
\_amcsouza@ufs.br\_}

\maketitle

\begin{history}
\received{Day Month Year}
\revised{Day Month Year}
\end{history}

\begin{abstract}
We investigate the classical Brownian motion of a particle in a
two-dimensional noncommutative (NC) space. Using the standard NC
algebra embodied by the sympletic Weyl-Moyal formalism we find that
noncommutativity induces a non-vanishing correlation between both
coordinates at different times. The effect stands
out as a
signature of spatial noncommutativity and thus
could offer a way to experimentally detect the phenomena. 
We further discuss some limiting scenarios and the 
trade-off between the scale imposed by the NC structure and the 
parameters of the Brownian 
motion itself. 

\end{abstract}

\ccode{PACS numbers:}


\section{Introduction}

How does the space-time structure look like as we gradually shift
towards smaller scales, say, the Planck scale? Is there some sort of
minimum length? Those are long-standing fundamental questions in
physics and has been the core of theories that attempt to join
gravity and quantum mechanics. In particular, there has been a
growing interest in investigating the role of noncommutative (NC)
geometries in nature \cite{douglas01}, namely when spatial
coordinates do not commute. This statement may sound a bit striking
as such property would limit our knowledge about the exact location
of a given particle in the space-time manifold, analogously to the
Heisenberg's uncertainty principle which imposes a fundamental limit
on the measurement precision of position and momentum variables.
Thereby, when assuming a NC algebraic framework, one adds a
minimum-length constraint into the problem.

The assumption that space-time is not continuous but, instead, a
quantized object goes back from Snyder's seminal paper
\cite{snyder47} where it was argued that a NC geometry
allows for regularized quantum field theories.
The overall idea was placed back again into the spotlight
by strong arguments from the string theory side a while ago \cite{connes98,
seiberg99}. In this way,
spatial noncommutativity indeed seems to play an essential role at the Planck's length
scale, where quantum effects of gravity might not be negligible
\cite{garay95, doplicher95, chamseddine96}.

Although noncommutativity embodies a puzzle piece in the high-energy
scenario \cite{carroll01, hewett01, chaichian05}, great interest has
also been addressed to its implications on condensed-matter physics
\cite{chaichian01, yu01, santos01, zhang04, ma11, santos12,
santos14}.
In particular,
the NC version of quantum mechanics \cite{chaichian01,gamboa01,ho02} has been
extensively explored over the past few years. Traces of noncommutativity have been studied, for
instance, in the hydrogen atom \cite{chaichian01, stern08, santos12,
santos14}, quantum Hall effect \cite{bellissard94, dulat01, ma11}, Aharonov-Bohm effect \cite{chaichian02, li06, liang14}, graphene \cite{santos01}, and even in quantum information theory \cite{bernardini13}.
Overall, the main motivation
turns out to be searching for observable signatures of
NC effects in more accessible
platforms, as well as setting experimental bounds on the NC scale
itself.
Interestingly, despite many proposals \cite{mathews01, falomir02, zhang04, zhang12, liang14}, there is still no actual
experimental evidence that holds the assumption of a NC
structure neither there is a way to prove it does not exist at all.
Here, we go along this direction and suggest 
another platform which can possibly enable 
its verification.
In particular, we take the
classical limit of NC quantum mechanics and deal with a well-known
model in statistical physics featuring a single particle going
through random displacements in a two-dimensional manifold, namely
the Brownian motion.

This model poses a fundamental importance in statistical physics by
describing the macroscopic picture of the particle due to
microscopic effects. This phenomena thus embodies how small-scale
physics can have a major influence at larger scales and then it
becomes natural to ask whether spatial noncommutativity plays any
significant role on it. Therefore, our aim is to place the
two-dimensional Brownian motion as a conceivable testbed to detect signatures of
spatial noncommutativity. Although dealing with 
the tiny scale where NC effects might emerge 
is physically demanding, the effect we describe here depends on many
properties of the Brownian motion itself which, in principle, 
could be manipulated so as to overcome this issue. 

It is worth mentioning that noncommutativity has also been explored
in the classical domain \cite{benczik02, romero03, romero03-kepler,
stern08, daszkiewicz08, vakili08, abreu12, abreu13}.
In order to do so, one can assume that the underlying algebra has a
sympletic structure compatible with that of NC quantum mechanics
\cite{chaichian01} (Dirac's correspondence principle). For instance,
a NC version of Newton's second law of motion was derived in
\cite{romero03}.

In the following, we set the theoretical ground for investigating
the Brownian motion in a two-dimensional NC manifold. This is done
by using the framework of NC algebra in the classical limit
\cite{romero03}. In particular, we solve the Langevin equation and
find that noncommutativity induces a time-dependent variance of the
correlation between spatial coordinates. 
For larger measurement-time differences, this quantity saturates 
to about a constant value that depends on the Brownian motion parameters such as
particle's size and density, temperature, and fluid viscosity. 
The key point here is that the
very fact of having such a non-zero correlation implies in the
existence of spatial noncommutativity.	

\section{NC formalism}

In NC quantum mechanics \cite{chaichian01}, the position
$\hat{x}_{i}$ and momentum $\hat{p}_{i}$ operators obey the
following commutation rules
\begin{eqnarray}
\left [\hat{x}_{i},\hat{x}_{j} \right ]&=&i\hbar\Theta_{ij},\cr
\left [\hat{x}_{i},\hat{p}_{j}\right ]&=&i{\hbar}\delta_{ij},\cr
\left [\hat{p}_{i},\hat{p}_{j}\right ]&=& 0,
\label{1.0}
\end{eqnarray}
with $i=1,2$, where $\hbar\Theta_{ij}$ is a antisymmetric matrix
with dimension of area and denotes the NC parameter.
Several studies have established
bounds on the scale of
$\Theta$ based on experimental data \cite{chaichian01, carroll01, falomir02,haghighat03, bertolami05, stern08,  ma11},
for instance, measurements of the
Lamb shift of the hydrogen atom gives
$\Theta \lesssim (6 \,\mathrm{GeV})^{-2}$ \cite{stern08}.

The mathematical framework for dealing with a NC space is
implemented via the Weyl-Moyal correspondence, where any arbitrary
function of the position operators $f(\hat{x})$ is associated with a
Weyl symbol $f(x)$ defined in the commutative scenario.
Hence, the usual product of two given
functions $f(\hat{x})g(\hat{x})$ is replaced by the so-called Weyl
star product $f(x)\star g(x)$ satisfying \cite{chaichian01}
\begin{eqnarray}
(f\star g)(x)=\mathrm{exp}\left(\frac{i}{2}\hbar\Theta_{ij}\partial_{i}\partial_{j}\right)f(x)g(y)\vert_{x=y},
\label{1.1}
\end{eqnarray}
where $f$ and $g$ are infinitely differentiable functions.

In the classical limit the commutators must be replaced with Poisson
brackets via the correspondence principle
\begin{eqnarray}
[\hat{A}, \hat{B}]\longrightarrow i\hbar\left\lbrace A, B \right\rbrace,
\label{1.2}
\end{eqnarray}
where $A$ and $B$ are two arbitrary functions. Thereby,
the relations in Eq. (\ref{1.0}) are rewritten as
\begin{eqnarray}
\left\lbrace x_{i}, x_{j} \right\rbrace &=&\Theta_{ij},\cr
\left\lbrace x_{i}, p_{j}\right\rbrace &=&\delta_{ij},\cr
\left\lbrace p_{i}, p_{j}\right\rbrace &=& 0.
\label{1.3}
\end{eqnarray}
Note that, in the classical limit, $\Theta$ must have
dimension of [time/mass] \cite{romero03-kepler, mirza04, djemai04}.

The general form of the Poisson brackets on the NC space are simply
worked out as
\begin{equation}
\left\lbrace A, B\right\rbrace = \left(\frac{\partial A}{\partial x_{i}}\frac{\partial B}{\partial p_{i}}-\frac{\partial A}{\partial p_{i}}\frac{\partial B}{\partial x_{i}} \right)+\Theta_{ij}\frac{\partial A}{\partial x_{i}}\frac{\partial B}{\partial x_{j}}.
\label{1.4}
\end{equation}
Now, consider the Hamiltonian
\begin{equation}
H = \dfrac{p_1^{2}+p_2^{2}}{2m}+V(x_1,x_2)
\end{equation}
describing a particle of mass $m$ in two dimensions subjected to an
external potential $V$. The equations of motion in the NC space are
then given by \cite{romero03}
\begin{equation}
\dot{x}_{i}=\left\lbrace x_{i}, H\right\rbrace =\frac{p_{i}}{m}+\Theta_{ij}\frac{\partial V}{\partial x_{j}},
\label{1.6}
\end{equation}
\begin{equation}
\dot{p}_{i}=\left\lbrace p_{i}, H\right\rbrace =-\frac{\partial V}{\partial x_{i}},
\label{1.7}
\end{equation}
and thus
\begin{eqnarray}
m\ddot{x}_{i}=-\frac{\partial V}{\partial x_{i}}+m\Theta_{ij}\frac{\partial^{2}V}{\partial x_{j}\partial x_{k}}\dot{x}_{k},
\label{1.8}
\end{eqnarray}
which represents Newton's second law in NC space
\cite{romero03}. Note that the spatial noncommutativity induces another force denoted by the
last term of the above equation. This
correction can be seen as an effective potential defined on the
NC background.

\section{Brownian motion}

Let us first introduce the Brownian motion in its standard
commutative version, usually described by the Langevin formalism
\cite{coffey04}. One can actually find many theoretical frameworks
to deal with it (see \cite{duplantier06} for a review). The Langevin
equation, however, stands out as a simple and straightforward
stochastic model which takes into account most the relevant physics
associated to the phenomena.
Let us consider a particle going through random displacements due to collisions with
(much smaller) molecules of a fluid and subjected to a viscous resistance force.
The Langevin equation accurately
describes the macroscopic dynamics of the Brownian particle in a
much longer time scale compared with the collision time
and is written as
\begin{eqnarray}
\frac{d\vec{v}(t)}{d
t}=-\frac{\gamma}{m}\vec{v}(t)+\frac{\vec{\xi}(t)}{m}, \label{1.9}
\end{eqnarray}
where $m$ is the particle's mass, $\gamma \vec{v}(t)$ denotes the viscous force
with coefficient $\gamma$, and $\vec{\xi}(t)$ is the noise
term forces arising from collisions in the fluid. The latter
satisfies the time-correlation functions
\begin{eqnarray}
\left\langle \xi_{i}(t)\right\rangle &=&0,\cr
 \left\langle \xi_{i}(t)\xi_{j}(t')\right\rangle &=&g \delta_{ij} \delta(t-t'),
\label{1.10}
\end{eqnarray}
where $g$ defines the force strength, $\delta_{ij}$
and $\delta(t-t')$ are, respectively, the Kronecker and Dirac delta
functions, and $\langle ... \rangle$ denotes the average with respect to
realizations of the random forces.
The correlations above define a second-order stochastic process
in which the random forces are given by a Gaussian distribution. Those
are completely uncorrelated at different times, thus
yielding a Markovian (memoryless) source of noise.

Now, let us move on to the NC algebra
set in Eq. (\ref{1.3}). Taking
$ \xi_{j}=-\partial V / \partial x_{j} $ as an external random force, from Eq. (\ref{1.6}) we get
\begin{eqnarray}
\frac{d x_{i}}{d t}=v_{i}(t)-\Theta_{ij}\xi_{j}(t).
\label{1.11}
\end{eqnarray}
Integrating Eq. (\ref{1.9}), we note that the solution for the particle's velocity
remains the same as in the usual commutative framework,
\begin{eqnarray}
v_{i}(t)=v_{0i}e^{-\frac{\gamma_{i}t}{m}}+\frac{1}{m}\int_{0}^{t}ds \; e^{-\frac{\gamma_{i}}{m}(t-s)}\xi_{i}(s),
\label{1.12}
\end{eqnarray}
where alongside Eq. (\ref{1.10}) the
expectation value of the quadratic velocity can be obtained:
\begin{eqnarray}
\left\langle v_{i}^{2}(t)\right\rangle =\frac{g}{2m\gamma}+\left(v_{0i}^{2}-\frac{g}{2m\gamma} \right)e^{-\frac{2\gamma}{m}t}.
\label{1.13}
\end{eqnarray}
In the long-time regime ($t \gg 1$), the above equation reduces to
\begin{eqnarray}
\left\langle v_{i}^{2}(t)\right\rangle =\frac{g}{2m\gamma},
\label{1.14}
\end{eqnarray}
thus yielding the so-called fluctuation-dissipation theorem
\cite{coffey04}.
For long times ($t \gg 1$), the system is driven towards a thermal equilibrium
state, balancing out fluctuation and dissipation effects. In this
scenario, the equipartition theorem becomes valid so that
\begin{eqnarray}
\frac{1}{2}m\left\langle v_{i}^{2}(t)\right\rangle_{eq}
=\frac{g}{4\gamma} =\frac{k_{\mathrm{B}}T}{2}, \label{21}
\end{eqnarray}
where $T$ denotes temperature and $k_B$ is the Boltzmann's
constant.

At this point, it is convenient to address
the diffusion coefficient, which can be extracted
from the particles's equations of motion.
Equation (\ref{1.11}) leads to
\begin{align}
x_{i}(t) &= x_{0i}+\frac{mv_{0i}}{\gamma}\left(1-e^{-\frac{\gamma}{m}t}\right) \nonumber \\
 & \;\; + \frac{1}{\gamma}\int_{0}^{t}ds\left( 1-e^{-\frac{\gamma}{m}(t-s)}\right)\xi_{i}(s)-\Theta_{ij}\int_{0}^{t}\xi_{j}(s)ds,
\label{1.15}
\end{align}
where along with Eq. (\ref{1.10}), the variance can be obtained,
\begin{align}
\sigma_{i}^{2}(t)&=\left\langle x_{i}^{2}(t)\right\rangle -\left\langle x_{i}(t)\right\rangle^{2} \nonumber \\
&= g\left(\frac{1}{\gamma^{2}}\!+\!\Theta_{ij}^{2} \right)t \!-\!
\frac{3mg}{2\gamma^{3}}\!-\! \frac{gm}{2\gamma^{3}}\left(
e^{-\frac{2\gamma}{m}t}\!-\!4e^{-\frac{\gamma}{m}t}\right).
\label{1.16}
\end{align}
In the long-time regime, the above expression turns into
\begin{eqnarray}
\sigma_{i}^{2}(t)=2D_{i}t-\frac{3mg}{2\gamma^{3}}, \label{1.17}
\end{eqnarray}
with
\begin{eqnarray}
D_{x}=D_{y}\equiv \frac{g}{2\gamma^{2}}\left(1+\Theta^{2}\gamma^{2}\right)
=
\frac{k_{\mathrm{B}}T}{\gamma}\left(1+\Theta^{2}\gamma^{2}\right)
\label{1.18}
\end{eqnarray}
being the diffusion coefficient along the $i$-axis (note that the
system is in thermal equilibrium). According to Eq. (\ref{1.18}), it
is clear that noncommutativity induces an extra term to the
diffusion coefficient, which is proportional to the square of
$\Theta$. If $\Theta=0$, we fully recover the usual commutative
description. It is worth to mention that this correction itself is
rather small to be detected experimentally and that is not what we want
to highlight. The most intriguing feature in considering the
Brownian motion in a NC manifold is shown in the following.

From Eqs. (\ref{1.10}) and (\ref{1.15}) we find that the NC nature
of space induces a non-vanishing variance of the correlation between
different coordinates at different times (say, $t_{1} > t_{2}$), in
contrast with the commutative case:
\begin{align}
\sigma_{xy} (t_{1},t_{2})&=\left\langle x(t_{1})y(t_{2})\right\rangle -\left\langle x(t_{1})\right\rangle\left\langle y(t_{2})\right\rangle  \nonumber \\
&= \frac{\Theta_{xy}gm}{\gamma^{2}}
\left(1-e^{-\frac{\gamma}{m}(t_{1}-t_{2})}\right)
+\frac{\Theta_{yx}mg}{\gamma^{2}}\left(e^{-\frac{\gamma}{m}t_{2}}-e^{-\frac{\gamma}{m}t_{1}}\right).
\label{1.19}
\end{align}
In thermal equilibrium ($t_{1} \gg 1$ and $t_{2} \gg 1$), and considering that $\frac{\gamma}{m}t_{1} \gg 1$ and $\frac{\gamma}{m}t_{2} \gg 1$, the above
equation reduces to  
\begin{eqnarray}
\sigma_{xy}(t_{1},t_{2}) = \frac{2m\Theta
k_{\mathrm{B}}T}{\gamma}\left(1-e^{-\frac{\gamma}{m}|t_{1}-t_{2}|}\right).
\label{1.20}
\end{eqnarray}
Note that we can, in a similar way, take $t_{2} > t_{1}$ [cf. Eqs. (\ref{1.10}) and (\ref{1.15})], what makes the above equation more general. 

We now address two limiting cases for that. First,
for $\frac{\gamma}{m}|t_{1} - t_{2}| \gg 1$, Eq. (\ref{1.20}) reads
\begin{eqnarray}
\sigma_{xy}(t_{1},t_{2}) = \frac{2m\Theta k_{\mathrm{B}}T}{\gamma}.
 \label{1.21}
\end{eqnarray}
On the other hand, for $ \frac{\gamma}{m} |t_{1} - t_{2}| \ll 1$ we have
(expanding the exponential and dropping out higher-order terms)
\begin{eqnarray}
\sigma_{xy}(t_{1},t_{2}) = 2\Theta k_{\mathrm{B}}T|t_{1}-t_{2}|.
\label{1.22}
\end{eqnarray}

\section{Discussion}

Equations (\ref{1.21}) and (\ref{1.22}) are the key results of this work.
The first thing we note is that spatial noncommutativity allows for the emergence
of correlations not seen in the standard commutative framework, as expected. Naturally, this very
scenario is recovered when $\Theta$ is null. 
Interestingly, the NC correction in Eq. (\ref{1.22}) features a time dependence
for short measurement-time differences, further saturating to Eq. (\ref{1.21}) 
for later times. 
In order to be able to make it physically attainable,
one must find a way to overcome the scale
imposed by $\Theta$. In our case, it implies in setting $\gamma/ m$
as low as possible. That would work for both 
regimes given by Eqs. (\ref{1.21}) and (\ref{1.22}). Most importantly,
the time difference in Eq. (\ref{1.22}) can (and ideally should) be made larger 
as long as we keep $|t_{1} - t_{2}| \ll m/\gamma$.
That would ultimately permit the observation of the time-dependence
of the variance.

In summary, the feasibility of probing NC effects
depends
upon the trade-off between 
the properties of Brownian particle along with its substrate and
the NC parameter. 
Considering a spherical Brownian particle with radius $a$ such that $m = \frac{4}{3}\pi a^{3}\rho$, with $\rho$ being the particle's density and using Stokes' formula $\gamma = 6\pi\eta a$ for a given fluid with viscosity $\eta$,
Eq. (\ref{1.21}) turns into
\begin{eqnarray}
\sigma_{xy}(t_{1},t_{2}) = \frac{4\Theta
k_{\mathrm{B}}T\rho a^{2}}{9\eta}.
 \label{1.23}
\end{eqnarray}
The above equation tells us that the particle's size and density
plays a significant role in setting the scale of the effect. Also,
it is desired to have very weak viscous forces acting on it.   
    
The Brownian motion is a well-established subject and
has been studied within various physical platforms with a high degree of precision and control \cite{haw02, lukic05, hanggi09, otsuka09, li10, huang11,duplat13,
millen14, kheifets14}. Still, there remains the challenge of
meeting the constraints imposed by the NC parameter. 
Nevertheless, the advantages of searching for signatures of
noncommutativity
on the space-time
structure of the Brownian motion are many: (i) it is an exactly solvable
model and finds a handful of applications; (ii) it is free of decoherence effects, unlike quantum
systems; (iii) recent advances in optical devices and
nanotechnology have increased the accuracy level in trajectory
analysis as well as in Brownian particle sizing \cite{lukic05,
hanggi09, otsuka09, li10, huang11,duplat13, millen14, kheifets14},
thus providing means to perform the experiment with a high degree
of resolution.


Another crucial aspect is that the Brownian motion shows self
similarity at any length and time scales thus establishing a valuable
platform to carry out studies on NC phenomena.
%
Furthermore, from a
theoretical point of view, it is highly
relevant to explore other
aspects of the motion itself,
e.g., its trace \cite{nienhuis82, kager04}.
A deep look at it could unveil solutions to bypass
the stringent range of parameters 
required to extract macroscopic signatures of noncommutativity. 
%
Our work further motivates the search for NC signatures in other
stochastic models, generalizing what we have found so far.

\section*{Acknowledgments}

This work was supported by CNPq (Grant No. 152722/2016-5) and CAPES (Grant No. 88881.030439/2013-01).
A.M.C.S. thanks the Instituto Nacional de
Ci\^{e}ncia e Tecnologia para Sistemas Complexos (INCT-SC).


\end{document}